\documentclass[emulateapj,epsfig]{article}
\usepackage{txfonts}
\usepackage{epsfig}
\usepackage{lscape}
\usepackage{graphicx}
\usepackage[onecolumn]{emulateapj}

\def\newline{\hfil\break}
\def\apj{{ApJ}}                 
\def\apjs{{ApJS}}               

\begin{document}
\title{The Evolution of Active Galactic Nuclei in Warm Dark Matter Cosmology}
\author{N. Menci, F. Fiore, A. Lamastra}
\affil{INAF - Osservatorio Astronomico di Roma, via di Frascati
33, I-00040 Monteporzio, Italy}
\smallskip
\vspace{0.cm}
\begin{abstract}
Recent measurements of the abundance of AGN with low-luminosities ($L_{2-10}\leq 10^{44}$ erg/s in the 2-10 keV energy band) at high redshifts $z\geq 4$ provide a serious challenge for Cold Dark Matter (CDM) models based on interaction-driven fueling of AGN. Using a semi-analytic model of galaxy formation we investigate how such observations fit in a Warm Dark Matter (WDM) scenario of galaxy formation, and compare the results with those obtained in the standard CDM scenario with different efficiencies for the stellar feedback. Taking on our previous exploration of galaxy formation in WDM cosmology, we assume as a reference case a spectrum which is suppressed - compared to the standard CDM case - below a cut-off scale $\approx 0.2$ Mpc corresponding (for thermal relic WDM particles) to a mass $m_X=0.75$ keV.  We run our fiducial semi-analytic model with such a WDM  spectrum to derive AGN luminosity functions from $z\approx 6$ to the present over a wide range of luminosities ($10^{43}\leq L_{2-10}$ /  erg s$^{-1}\leq 10^{46}$ in the 2-10 keV X-ray band), to compare with recent observations and with the results in the CDM case. When compared with the standard CDM case, the luminosity distributions we obtain assuming a WDM  spectrum are characterized by a similar behaviour at low redshift, and by a flatter slope at faint magnitudes for $z\geq 3$, which provide an excellent fit to present observations. We discuss how such a result compares with CDM models with maximized feedback efficiency, and how future deep AGN surveys will allow for a better discrimination between feedback and cosmological effects 
on the evolution of AGN in interaction-driven models for AGN fueling.
\end{abstract}
\vspace{-0.2cm}
\section{Introduction}
Perhaps the most striking aspects of Active Galactic Nuclei (AGN) are their strong evolution with redshift, and the strong correlations between the properties of  the AGN and those of their host galaxies. The former appears as a strong increase in their number and luminosities from $z=0$ to $z=3$ (see Hartwick \& Shade 1990; Boyle et al. 2000; Ueda et al. 2003) followed by a decline in the number of AGN at higher redshifts (Fan et al. 2001, 2004),  while the latter  involve, e.g.,  the correlations between the stellar mass (or  the velocity dispersion of the bulge) of the host galaxy and the mass of the accreting Black Hole powering the AGN emission  (Kormendy \& Richstone 1995; Magorrian et al. 1998; Ho 1999; Gebhardt et al. 2000; Ferrarese \& Merritt 2000; Marconi \& Hunt 2003; H\"aring \& Rix 2004; Kormendy \& Bender 2009). Both such key features call for a cosmological explanation of the Black Hole and AGN properties, which in fact constitutes one of the most pressing issues in astrophysics. 

On the theoretical side, models aimed at describing the evolution of Black Holes and AGN in a cosmological context are based on either N-body simulations and semi-analytic models. While the former  have shown the importance of galaxy mergers as triggers for AGN accretion  and the role of the AGN energy feedback in the subsequent evolution of the host galaxy (see, e.g., Di Matteo, Springel \& Hernquist 2005, Springel 2005; Hopkins 2005a, 2006), the evolution of the statistical properties of AGN in a cosmological context has been mainly investigated through SAM (see Baugh 2006 for a review), which  allow a faster spanning of the parameter space compared to N-body simulations. In fact, they adopt analytic laws to connect the evolution of Dark Matter (DM) haloes collapsed from the primordial density field and their subsequent merging histories (derived from Monte Carlo or N-body simulations) to the physics of baryons inside the haloes, including the gas processes, the star formation and the growth of supermassive Black Holes (BHs) due to gas accretion and merging (see Kauffmann \& Haehnelt 2000; Menci et al. 2003, 2006; Croton et al. 2006; Bower et al. 2006; Cattaneo et al. 2006; Hopkins et al. 2006; Monaco, Fontanot \& Taffoni 2007; Marulli et al. 2008). 

Beside some variance concerning the role of minor interactions and the implementation of the AGN feedback, quenching the star formation in the host galaxy, all such models include galaxy interactions as triggers for gas accretion onto the BHs; observationally, this is motivated by the ULRIG-QSO connection (Sanders \& Mirabel 1996, Canalizo \& Stockton 2001), by the signature of recent mergeres in QSO hosts (see, e.g., Bennert et al. 2008; Treister et al. 2012), and the smal-scale over densities around luminous QSOs (see Fisher et al. 1996; Bahcall et al. 1997; Serber et al. 2006; Hennawi et al. 2006; Myers et al. 2008; Strand et al. 2008). Based on the assumption of interaction-triggered QSOs, complemented with the inclusion of accretion of hot gas (Bower et al. 2006; Croton et al. 2006) and of disk instabilities as additional triggers (see Fanidakis et al. 2012; Hirschmann et al. 2012), 
several state-of the-art cosmological models correctly describe the tight correlation between galaxy properties and black hole  mass (Hopkins et al. 2008; Marulli et al. 2008; Somerville et al. 2008, Lamastra et al. 2010; Guo et al. 2011) and the evolution of luminous QSO population over a wide range of cosmic ages (see Menci et al. 2003, 2006; Hopkins et al. 2008; Shankar et al. 2010;  Fanidakis et al. 2012; Hirschmann et al. 2012). These successes allowed the models to provide a cosmological interpretation of the observed luminosity-dependent evolution of AGN (the  "downsizing" effect):
on the one hand, the strong evolution of the bright QSO population observed in optical surveys (see Richards et al. 2006; Croom et al. 2009; Glikman et al. 2011; Jiang et al. 2009) results from the combination of the time-decline of merging  activity and of the exhaustion of the cold gas content in massive galaxy haloes due to the its early conversion into stars at high redshifts; on the other hand, the smoother evolution with redshift of the observed abundance of low-luminosity AGN revealed mainly by X-ray surveys (effectively probing such a population, see e.g. Hasinger, Miyaji, Schmidt  2005; La Franca et al. 2005;  Fiore et al. 2012 and discussion therein) results from the milder evolution of the gas content in low-mass galaxies predicted by the models and the early collapse epoch of the corresponding host DM clumps envisaged in the standard CDM cosmology. 

However, while for bright QSOs and for low-redshift AGN the predicted evolution and downsizing agree with measurements from optical and X-ray surveys (see Menci et al. 2008; Hopkins et al. 2008; Shen  2009; Shankar et al. 2010), the recent observational breakthroughs concerning the abundance of faint AGN at at high redshifts $z\gtrsim 3-5$ are posing serious challenges to the cosmological models for the AGN evolution. 
Indeed, recent estimates of the X-ray luminosity functions of AGN extending down to log $L$(2 -10 keV)$\approx 42.75$ (Fiore et al. 2012) 
are in sharp contrast with the large density of faint AGN predicted by analytic (Shankar et al. 2010) or semi-analytic (Menci et al. 2006, 2008) models  for the co-evolution of galaxies and AGN (see also Shankar \& Mathur 2007; Wyithe \& Loeb 2003; Lapi et al. 2006), even when the observed luminosity functions are corrected for the estimated fraction of obscured objects (e.g., La Franca et al. 2005), the models 
predicting up to ten times more AGN with X-ray luminosity $L_X\leq 10^{43}$ erg/s (in the 2-10 keV band). 
Results from other last-generation semi-analytic models based on merging trees extracted from N-body simulations (Marulli et al. 2008) or on Monte Carlo generated merging trees (Hirschmann et al. 2012) confirm the mismatch. In fact, although the above papers also assume disk instabilities to provide additional triggers, they yield high-redshifts ($z\gtrsim 4$) AGN luminosity functions steeper than the observed ones, so that they cannot provide a simultaneous fit to both the number density of bright QSOs ( measured by the SLOAN survey) and the abundance of lower luminosity AGNs (detected by from hard X-ray measurements), unless tuned, non-canonical assumptions (like heavy BH seeds, a varying sub-Eddington limit, and a halo mass limit) are adopted (Hirschmann et al. 2012). 
The indication is that, when tuned to account for the observed high abundance of low-redshift faint AGNs and of bright QSO at high redshifts, most semi-analytic models based on CDM are characterized by an over prediction of faint AGNs at $z\gtrsim 4$.  Note however that the possibility that the CDM models of AGN evolution may be brought in agreement with the observations 
by a proper implementation of the baryon physics is still open; in fact, the AGN luminosity functions derived in the CDM semi-analytic model by Fanidakis et al. (2012) have a flatter slope at the faint end as to be consistent with the observations (including those concerning  the X-ray luminosity distributions after their modelling of the fraction of obscured objects). 

Such an over prediction of faint objects, especially at high redshifts, is indeed  a common and robust feature of CDM galaxy formation models. In fact,  the latter yield galaxy luminosity function largely exceeding those measured in K-band  for $0\lesssim z\lesssim 3$  (see Cirasuolo et al. 2010; Henriques et al. 2011) or those corresponding to faint Lyman-break galaxies (Lo Faro et al. 2009) at $z\gtrsim 3$; the excess of star-forming low-mass galaxies at $z\gtrsim 3$ reflects into an excess of red low-mass galaxies at $z\approx 0$ (see, e.g., Croton et al. 2006; Salimbeni et al. 2008). Analogously, the number of galaxies with stellar mass $M_*\lesssim 10^{10}$ M$_{\odot}$  in the Universe is systematically over predicted by
all theoretical CDM models (Fontana et al. 2006; Fontanot et al. 2009; Marchesini et al. 2009; Guo et al. 2011) in the whole  range $1\lesssim z\lesssim 3$ where the small-mass end of the mass function has been measured (see also Santini et al. 2011).

While it is possible that AGN feeding mechanisms not based on galaxy interaction (see Ciotti 2009; Shin et al. 2010; see also Treister et al. 2012 for 
the relative role of merging and secular processes) may be at the origin of the mismatch, the dominating point of view is that all models miss some kind of process involved in the complex baryonic physics describing galaxy formation and AGN evolution (see, e.g., Somerville et al. 2008). In particular, models could fail to properly describe the 
feedback processes such as heating or winds caused by Supernovae explosions and UV background, which may be effective in  expelling gas in shallow potential wells thus suppressing both star formation and BH growth in low-mass galaxies. In present models the effectiveness of such a solution is limited by the large densities of DM haloes formed at high redshift, which allow for an effective shielding of the inner regions and 
yield escape velocities larger than the kinetic energy of SN driven winds (see discussion in Lo Faro et al. (2009). 

An alternative possibility is that the conflicts are rooted in the CDM spectrum assumed by all above cosmological models of galaxy formation and AGN evolution. Indeed, the shape of the CDM spectrum of density perturbations, ultimately determining the abundance and the merging histories of DM haloes, 
is  constrained by present observations only for mass scales exceeding $M\approx 10^{9}\,M_{\odot}$, corresponding to scales $r\gtrsim 0.3$ Mpc effectively probed by the comparison of the Lyman-$\alpha$ forest spectra at $z\approx 2.5$ with the N-body simulations (Viel et al. 2005; 2008). 
On smaller mass scales, a strong suppression in the DM spectrum would still be consistent with present observations; physically, such a suppression 
could be achieved by assuming DM to be constituted by  particles with mass $m_X\approx 1$ keV, characterized by large thermal speeds $v_s/c\approx 0.2$ (Warm Dark Matter, WDM, see Peebles 1982; Bond, Szalay \& Turner 1982; Colombi, Dodelson, Lawrence 1996 and references therein; see also De Vega \& Sanchez 2012) corresponding to a free-streaming scale $r_{fs}\sim v_s\,t_{eq}\approx 0.2$ Mpc  (here $t_{eq}$ is the time of matter-radiation equality).
An effective suppression of the DM power spectrum at small scales could indeed provide a viable solution to several long-standing tensions between the CDM predictions and the observations, like  the low number of satellites within the haloes of galaxies and 
groups compared to CDM model predictions (see Klypin et al. 1999, Moore et al. 1999; see also Mateo 1998; Lovell et al. 2012) and the observed inner density profiles of galaxies inferred from rotation curves of real galaxies which are flatter than those resulting in CDM N-body simulations (Moore et al. 1999, Abadi et al. 2003, Reed et al. 2005; Madau, Diemand, Kuhlen 2008; for observations see, e.g., Gentile et al. 2007, McGaugh et al. 2007). 

In a previous paper (Menci et al. 2012, Paper I) we investigated for the first time the effects of a WDM power spectrum on the statistical properties of galaxies using our  semi-analytic model of galaxy formation, and we showed that assuming WDM cold alleviate the discrepancy between the observed and the  predicted galaxy luminosity and stellar mass distributions.  Here we take on our previous study to explore the evolution of AGN in a WDM context; we focus on the evolution of the AGN luminosity 
function from the highest redshift where observations are available ($z\approx 6$) to the present, and we compare with state-of-the-art observational data and with our previous  predictions based on the standard CDM cosmology. 
 \newpage

 \section{Method}
 We take on the method adopted in Paper I to investigate the effects of the WDM power spectrum on galaxy formation.
We adopt the Rome semi-analytic model (R-SAM) to connect the physical processes involving baryons 
(physics of gas, star formation, feedback, growth of supermassive Black Holes) to the merging histories
of DM haloes, ultimately determined by the DM initial power spectrum. The model free parameters, including the cosmological ones,  are set as in 
Menci et al. (2005, 2006, 2008); the only change is that merging trees are now computed in a WDM cosmology. This allows to single out the effects of changing the DM spectrum with the same  
baryon physics and cosmological framework. Here we briefly recall the basic features of the model (for a complete description see Menci et al. 2005, 2006, 2008), 
and we describe the power spectrum we use to compute the merging trees in the WDM cosmology. 
\subsection{The Semi-Analytic Model: The History of Dark Matter Haloes in CDM and WDM Cosmology}

Galaxy formation and evolution is driven by the collapse and growth of DM haloes, which originate 
from the gravitational instability of  overdense
regions in the primordial density field. This is taken to be a random,
Gaussian  density field within the
''concordance cosmology" (Spergel et al. 2007), for which we adopt round
parameters  $\Omega_{\Lambda}=0.7$, total matter density parameter $\Omega_{M}=0.3$ 
(with baryons contribution corresponding to  $\Omega_b=0.04$ and Hubble constant (in units of 100 km/s/Mpc) $h=0.7$. The
normalization of the spectrum is taken to be $\sigma_8=0.9$ in terms of the variance
of the field smoothed over regions of 8 $h^{-1}$ Mpc. Adopting the exact best fit 
WMAP7 values for the above parameters does not change our results appreciably. 

As  cosmic time increases, larger and larger regions of the density field
collapse, and  eventually lead to the formation of groups and clusters of
galaxies; previously formed, galactic size  condensations are enclosed. 
The statistical properties and the merging history of DM haloes depend on the variance 
of the primordial DM density field as a function of the smoothing mass scale: 
\begin{equation} 
\sigma^2(M)=\int {dk\,k^2\over 2\,\pi^2}\,P(k)\,W(kr)~,
\end{equation}
where $P(k)$ is the linear power spectrum of DM perturbations at a wavelengths 
$k=2\pi/r$, $M\simeq 1.2\,10^{12}\,h^2\,{\rm M}_{\odot}\,(r/{\rm Mpc})^3$ is the the mass within a sphere 
of radius $r$, and $W$ is a window function (see 
Peebles 1993), usually assumed to be top-hat in real space. 
Thus, the linear power spectrum $P(k)$ determines the merging histories and the mass distribution of DM haloes. For the CDM cosmology 
we adopt the form $P_{CDM}(k)$ given by Bardeen et al. (1986). The WDM spectrum is suppressed with respect to the CDM case below a characteristic scale depending on the mass of the WDM particles (and, for non-thermal particles, also on their mode of production, see Kusenko 2009); in fact, the large thermal velocities of the lighter WDM particles erase the perturbations with size comparable and below the free-streaming scale $r_{fs}$. 
In the case WDM is composed by relic thermalized particles, such a  suppression  is quantified 
through the ratio of the two linear power spectra (transfer function) which can be parametrized as (Bode, Ostriker \& Turok 2001, see also Viel et al. 2005)
\begin{equation}
{P_{WDM}(k)\over P_{CDM}(k)}=\Big[1+(\alpha\,k)^{2\,\mu}\Big]^{-5\,\mu}~~~~~~~~{\rm with} ~~~~~~~~\alpha=0.049 \,
\Big[{\Omega_X\over 0.25}\Big]^{0.11}\,
\Big[{m_X\over {\rm keV}}\Big]^{-1.11}\,
\Big[{h\over 0.7}\Big]^{1.22}\,h^{-1}\,{\rm Mpc}~~~~~~~~{\rm and}~~~~~~~\mu=1.12
\end{equation}
The above relation between transfer function and DM particle mass $m_X$ is valid for thermal relics; a similar relation 
holds for sterile neutrinos (if they are produced from oscillations with active neutrinos) provided one substitutes the mass $m_X$ with a mass \\$m_{sterile}=4.43\,{\rm KeV}\,
(m_X/{\rm keV})^{4/3}\,(\Omega_{WDM}\,h^2/0.1225)^{-1/3}$ . In both cases, the smaller the WDM mass $m_X$ (or $m_{sterile}$) the larger the suppression with respect to the standard CDM spectrum. We shall adopt as a reference case a  WDM (thermal) particle mass $m_X=0.75$ keV; this allows us to investigate the effects of the  largest  suppression of the DM power spectrum which is still consistent with the most stringent observational constraints $m_X\gtrsim 0.6$ keV (for thermal particles); these have been derived by Viel et al. (2005) by comparing the observed Lyman-$\alpha$ forest in absorption spectra of Quasars at $z=2-3$ with the results of N-body simulations run assuming different WDM  power spectra of perturbations. 
Adopting the above value for the WDM particle mass yields a suppression of the power spectrum (with respect to CDM) at scales below $r_{fs}\approx 0.2$ Mpc (see Paper I); correspondingly, the abundance and the probability of inclusion of DM haloes with mass $M_{fs}\approx 5\,10^{8}\,M_{\odot}$ are suppressed (with respect to the CDM case), according to eq. 1. 

The  mass function and the progenitor distribution of DM haloes (determining the merging history of the haloes)  can be derived from eqs. 1 and 2 (see Barkana et al. 2001; Smith \&  Markovic 2011), assuming a threshold for collapse of DM haloes given by the critical linear-theory overdensity fo the collapse of top-hat perturbations. For the CDM case, a detailed framework for the computation of such quantities (the Extended Press \& Schechter formalism, EPS hereafter) has been introduced by Bond et al. (1991; see also Lacey \& Cole 1993), and later developed to improve the agreement with N-body simulations (see Sheth , Mo \& Tormen 2001).  For the WDM case, recent works (Benson et al. 2013) have shown that the choice of an appropriate window function $W$ (eq. 1) and of the collapse threshold is non-trivial; the latter authors provide a generalized algorithm for computing the halo merger rates and the  mass function in WDM cosmology where the threshold and the filter function are calibrated through N-body simulation. They found that, compared with a straightforward application of EPS with a WDM spectrum (eq. 2), both the progenitor distribution and the mass function are strongly suppressed below the free-streaming scale $M_{fs}$. In our Paper I we assumed the straightforward extension  of the EPS to the WDM spectrum as a reference framework to compute the above quantities, and we bracketed the uncertainties concerning the window function and the collapse threshold by assuming a sharp cutoff  for the progenitor distribution and the mass function at $M=M_{fs}$, a phenomenological (extreme) rendition of the findings of Benson et al. (2013). Here we retain such approach, and we shall show the WDM results on the AGN luminosity functions as bracketed between those corresponding to a straightforward application of the EPS and those corresponding to a merging history where the progenitor distributions (and correspondingly the halo mass functions) are completely suppressed for masses $M\leq M_{fs}$; the detailed procedure adopted to generate Monte Carlo realizations of merger trees in both cases is described Sect. 4 of Paper I.

Following the canonical procedure adopted by SAMs, after generating Monte Carlo realizations the DM merger trees we follow 
the history of DM clumps included into larger DM haloes; these may survive as satellites, or merge to form larger galaxies due to binary
aggregations,  or coalesce into the central dominant galaxy due to dynamical friction (see Menci et al. 2005, 2006); in addition satellite halos are partially disrupted as the density in their outer parts becomes lower than the density of the host halo within the pericentre 
of its orbit (see Menci et al. 2002 for details). Since the above galaxy coalescence 
processes take place over time scales that grow longer over cosmic time,  the number of satellite galaxies increases as the DM host haloes
grow from groups to clusters. 

\subsection{The Semi-Analytic Model: The Baryonic Physics and Evolution of AGN}

The SAMs relate the physics of baryons to the 
DM merging trees, determined by the DM power spectrum as discussed above. 

The radiative gas cooling, the ensuing star formation and the
Supernova events with the associated feedback occurring  in the growing
DM haloes are computed for each sub-halo hosting a galaxy. 
The cooled gas with mass $m_c$ settles into a rotationally supported disk with radius $r_d$, rotation velocity $v_d$
and dynamical time $t_d=r_d/v_d$, all related to the DM sub-halo mass. The gas gradually condenses  into
stars at a rate $\dot m_*\propto m_c/t_d$. The stellar ensuing feedback
returns part of the cooled gas to the hot gas phase
at the virial temperature of the halo; from a simple energy balance argument, the mass rate of the  gas returning to the hot phase can be estimated as $\dot m_{hot}=\epsilon_0\eta_{SN}\,E_{SN}/v_{esc}^2$ (Dekel \&Silk 1986; cf. Kauffmann et al. 1993; Natarajan 1999), where $v_{esc}$ is the galaxy escape velocity, $E_{SN}=10^{51}$ erg is the energy released by a single Supernova, $\eta_{SN}$ is the number of Supernovae per unit stellar mass (for a Salpeter IMF $\eta_{SN}= 6.5\,10^{-3}$) and $\epsilon_0\in [0,1]$ is a tunable efficiency for the coupling of the emitted energy with the interstellar gas: in our fiducial model we set $\epsilon=0.01$. An additional
channel for star formation implemented in the model is provided by
interaction-driven starbursts, triggered not only by merging but
also by fly-by events between galaxies; such a star formation mode
provides an important contribution to the early formation of stars
in massive galaxies, as described in detail in Menci et al. (2003,
2005). 

The R-SAM provides a detailed description of the growth of Black Holes (BHs) inside each DM halo from primordial seeds 
with mass $M_{BH}\approx 10^2\,M_{\odot}$ through merging between galaxies and 
accretion of a fraction  of the cold galactic gas $m_c$. The accretion is triggered by 
galaxy  interactions (i.e., minor and major merging, and fly-by events), and the fraction of cold gas funnelled to the central BH 
is derived from the model by Cavaliere \& Vittorini (2000). The active accretion phase of BHs
corresponds to the AGN. 

The accretion of cold gas is triggered by galaxy interactions, namely, fly-by encounters, and
both minor and major merging events. In fact, all such kinds of interactions destabilize part of the cold gas available 
by inducing loss of angular momentum. In fact, small scale (0.1-a few $kpc$) regions are likely to have 
disk geometry if they are to efficiently remove angular momentum and convey to 
the (unresolved) pc scales the gas provided on larger scales (these 
may be isotropized by head-on, major merging events). 
The fraction of cold gas accreted by the BH in an interaction event is computed
in terms of the variation $\Delta j$ of the specific angular momentum $j\approx
Gm/v_d$ of the gas, to read (Menci et al. 2003) 
\begin{equation}
f_{acc}=10^{-1}\Big\langle {m'\over m}\,{r_d\over b}\,{v_d\over V}\Big\rangle\, .
\end{equation}
Here $b$ is the impact parameter (evaluated as the average distance of the
galaxies in the halo),  $m'$ is the mass of the  partner galaxy in the
interaction,  and the average runs over the probability of finding such a galaxy
in the same host halo (with circular velocity $V$) where the galaxy with mass $m$ is located.
The values of the quantities involved in the average yield values of $f_{acc}\lesssim 10^{-2}$. 
For minor merging events and for  fly-by encounters among galaxies with very unequal mass ratios
$m'\ll m$, dominating the statistics in all hierarchical models of galaxy formation, the 
accreted fraction takes values $10^{-3}\lesssim f_{acc}\lesssim 10^{-2}$. 
The average amount of cold gas accreted during an accretion episode is thus
$\Delta m_{acc}=f_{acc}\,m_c$, and the duration of an accretion episode, i.e.,
the timescale for the QSO or AGN to shine, is assumed to be the crossing time
$\tau=r_d/v_d$ for the destabilized cold gas component.

The rate of such  interactions is given by Menci et
al. (2003) in the form
$\tau_r^{-1}=n_T\,\Sigma (m,m')\,V_{rel}.$
Here $n_T$ is the number density of galaxies in the same host halo and
$V_{rel}=\sqrt{2}\,V$ is their average relative velocity in the host halo with 
circular velocity $V$. The cross section $\Sigma$ for grazing 
encounters ( effective for angular momentum transfer) is given by Saslaw (1985) in terms of the tidal radii  
associated to the two interacting partners with mass $m$ and $m'$ 
(see Menci et al. 2003, 2004). For each galaxy (with mass $m$) in the Monte Carlo realizations, and 
at each time step $\Delta t$, the probability 
to interact with a partner (with mass $m'$) in the same halo is estimated as $\Delta t/\tau_r(m,m')$. 

For each galaxy during an interaction phase, the  time-averaged bolometric luminosity so produced by a QSO hosted in a given galaxy 
is then computed  as
\begin{equation}
L={\eta\,c^2\Delta m_{acc}\over \tau} ~.
\end{equation}
We adopt an energy-conversion efficiency $\eta= 0.1$ (see Yu \&
Tremaine 2002), and derive luminosities in the various bands adopting 
standard spectral energy distributions following Marconi et al.
(2004) in the X-ray and UV bands. The supermassive BH mass $m_{BH}$ grows mainly through accretion
episodes as described above, besides  coalescence with other BHs
during galaxy merging. As initial condition, we assume small seed
BHs of mass $10^2\,M_{\odot}$ (Madau \& Rees 2001) to be initially
present in all galaxy progenitors; our results are insensitive to
the specific value as long as it is smaller than some
$10^5\,M_{\odot}$.

The radiation energy released by AGN into the interstellar medium  heat and expel part of the galactic gas 
(Cavaliere, Lapi \& Menci, 2002), as indicated, e.g., by the fast winds with velocities up to $10^{-1}c$  observed in the central regions of AGN, likely originating from the acceleration of disk outflows by the AGN radiation field (see Begelman 2003 for a review). 
A detailed model for the transport of energy from the inner, outflow region to the larger scales has been developed by  Lapi et al. (2005), and implemented into the R-SAM by Menci et al. (2008). Central, highly supersonic outflows compress the gas into a blast wave terminated by a leading shock front, which moves outwards with a lower but still supersonic speed and sweeps out the surrounding medium. 
The key quantity determining all shock properties is the total energy injected by AGN into the surrounding gas $\Delta E=\epsilon _{AGN} \eta c^2 \Delta m_{acc}=\epsilon _{AGN}\,L\,\tau$. 
 This is computed for each BH accretion episode in our Monte Carlo simulation; the value of the energy feedback efficiency for coupling with the surrounding gas is taken as $\epsilon _{AGN}=5\times 10^{-2}$, which is consistent with the values required to match the X-ray properties of the ICM in clusters of galaxies (see Cavaliere et al. 2002). This is also consistent with the observations of wind speeds up to $v_w\sim 0.1c$ in the central regions, which yield $\epsilon _{AGN} \sim v_w/2c\sim 0.05$ by momentum conservation between photons and particles (see Chartas et al. 2002; Pounds et al. 2003); this value has also been adopted in a number of simulations (e.g., Di Matteo et al. 2005, Sijiacki et al. 2007) and semianalytic models of galaxy formation (e.g., Menci et al. 2006, see Hopkins 2005).
The  injection of energy $\Delta E$ (relative to the initial thermal energy content 
$E\propto m_c$ of the galactic gas) determines the Mach number  ${\mathcal M}\approx (1+\Delta E / E)^{1/2}$ (Lapi et al. 2005)
of the expanding shock sweeping out the galactic gas  surrounding the AGN. Thus, the expansion velocity of the bast wave, and hence the properties of the surrounding galactic gas,  are directly related to the AGN luminosity. Note that in principle the same AGN feedback affects not only the galactic gas but also the halo hot gas. However, the key quantity in our treatment of the AGN feedback is the ratio $\Delta E/E$; since the thermal energy content of the hot halo gas is much larger than that of a single galaxy, the effect on the hot gas is smaller than that on the galaxies and does not affect our main results below. 

The above modelling of AGN evolution in the framework of hierarchically evolving DM haloes has been worked out in our previous papers to compare with a wide set of observations. The predicted local $M_{BH}-m_*$ relation has been successfully tested against observations in Lamastra et al. (2010); the description of AGN feedback yields an inverse luminosity dependence of the fraction of obscured AGN (with column densities $N_H\geq 10^{22}$ cm$^{-2}$) which correctly reproduces the observed behaviour (Menci et al. 2008); finally, the predicted abundance of high-luminosity AGN agrees with the observed luminosity functions up to the highest redshifts $z\approx 6$ (see Menci et al. 2004; 2008). However, as discussed in the Introduction, the abundance of low-luminosity AGN withX-ray luminosity $L_X\leq 10^{43}$ erg s$^{-1}$ at high redshifts $z\geq 2$ is largely over-predicted by the model, independently of the tuning of free parameters (e.g.,  the feedback efficiency $\epsilon_{AGN}$), at least within the ranges allowed by consistency of the AGN and host galaxy properties with independent observables. 
In the following, we shall investigate the impact of a WDM spectrum on such a  population of low-luminosity AGN. 
\vspace{0cm}
\section{The evolution of the AGN luminosity functions in a WDM cosmology}
To compute the evolution of the AGN luminosity function in the CDM and WDM, we first run Monte Carlo realizations of merging trees in both cosmologies as discussed in Sect. 2.1. Then we run the full SAM to model the baryonic physics in the generated DM halos, as described in Sect. 2.2; the model free parameters are the same as in our previous papers (Menci et al. 2005, 2006, 2008, 2012) in order to 
single out the effects of changing the DM spectrum with the same baryon physics. 

We show the resulting evolution of the AGN luminosity function in fig. 1. Since we aim at investigating the effects of suppressing the power spectrum at small masses (hence at low AGN luminosities), we chose to show the distribution of X-ray luminosities in the 2-10 keV band; 
this allows us to directly compare with X-ray observations, which are best suited to sample the population of low-luminosity AGN. 
The predictions for the WDM case are represented by shaded areas, with the upper envelope corresponding to merger trees derived by directly extending the standard EPS theory to the WDM case, while the lower envelope of the area corresponds to assuming a sharp cutoff for the progenitor distributions (and halo mass functions) for $M\leq M_{fs}$ (see sect. 2.1 and sect. 4 in Paper I). 

While at low redshifts $z\lesssim 1.5$ both the CDM and the WDM models provide acceptable fits to the observations (with WDM slightly underestimating the abundance of low luminosity AGN), at higher redshifts large differences arise at luminosities $L_{2-10 keV}\lesssim 10^{44}$ erg/s. At such redshifts and luminosities, the CDM model yields luminosity functions appreciably steeper than observed; this behaviour of CDM predictions is also shared by other recent semi-analytic models for the evolution of AGN (Hirschmann et al. 2012; see also Shankar et al. 2010), so that it seems to constitute a  common feature of most fiducial CDM models, although radically different assumptions like heavy initial BH seeds (Begelman, Volonteri, Rees 2006; see also Khlopov, Rubin, Sakharov 2004) could still resolve the mismatch (see the Conclusions). A better agreement is obtained in the WDM case, due to the suppression of the power spectrum at small scales (eq. 2). This  reduces the predicted density of low-luminosity AGN since - in an interaction-driven model -  low luminosity AGN are related either to low-mass galaxies accreting clumps  of comparable mass (low values of the cold gas mass $m_c$, see text between eqs. 4-5), or to larger galaxies interacting with clumps of much smaller mass (the $m'/m$ ratio in $f_{acc}$ in eq. 4). In both cases, reducing the abundance of low-mass condensations through the adoption of a WDM model results into a flatter AGN luminosity function at the low-mass end. The effect is stronger at higher redshifts since CDM models predict low-mass galaxies to form mainly at early cosmic times, so that at $z\lesssim 2$ a substantial fraction of them has been included into larger galaxies; on the other hand, the (minor) fraction of low-mass galaxies collapsed at $z\lesssim 2$ is characterized by a lower gas densities (corresponding to lower escape velocities) and hence by a larger gas depletion due to Supernovae feedback, which contributes to flatten the AGN luminosity function even in the CDM case. 
We expect the above effects to be quite general in interaction-driven scenarios for AGN fueling, since the $m'/m$ scaling of the 
destabilized gas (and of the related BH accretion, see, e.g., Hopkins and Quataert 2010) results not only from our model (see eq. 3), but also from aimed hydrodynamical N-body simulations
of galaxy merging (see Cox et al. 2008), while the direct scaling of $m_c$ with the galaxy mass is shared by all cosmological models of galaxy formation, as well as the feedback dependence on the escape velocity of galaxies. Note also that - at any redshift -  for the luminous QSO population the effect of assuming a WDM spectrum is minimal. In fact, the large accretion rates characterizing such objects require the interactions of massive  galaxies with clumps of comparable sizes, whose abundance is not sensibly affected by the small-scale cutoff in the DM power spectrum.

\begin{center}
\vspace{-0.4cm}
\scalebox{0.64}[0.64]{\rotatebox{-90}{\includegraphics{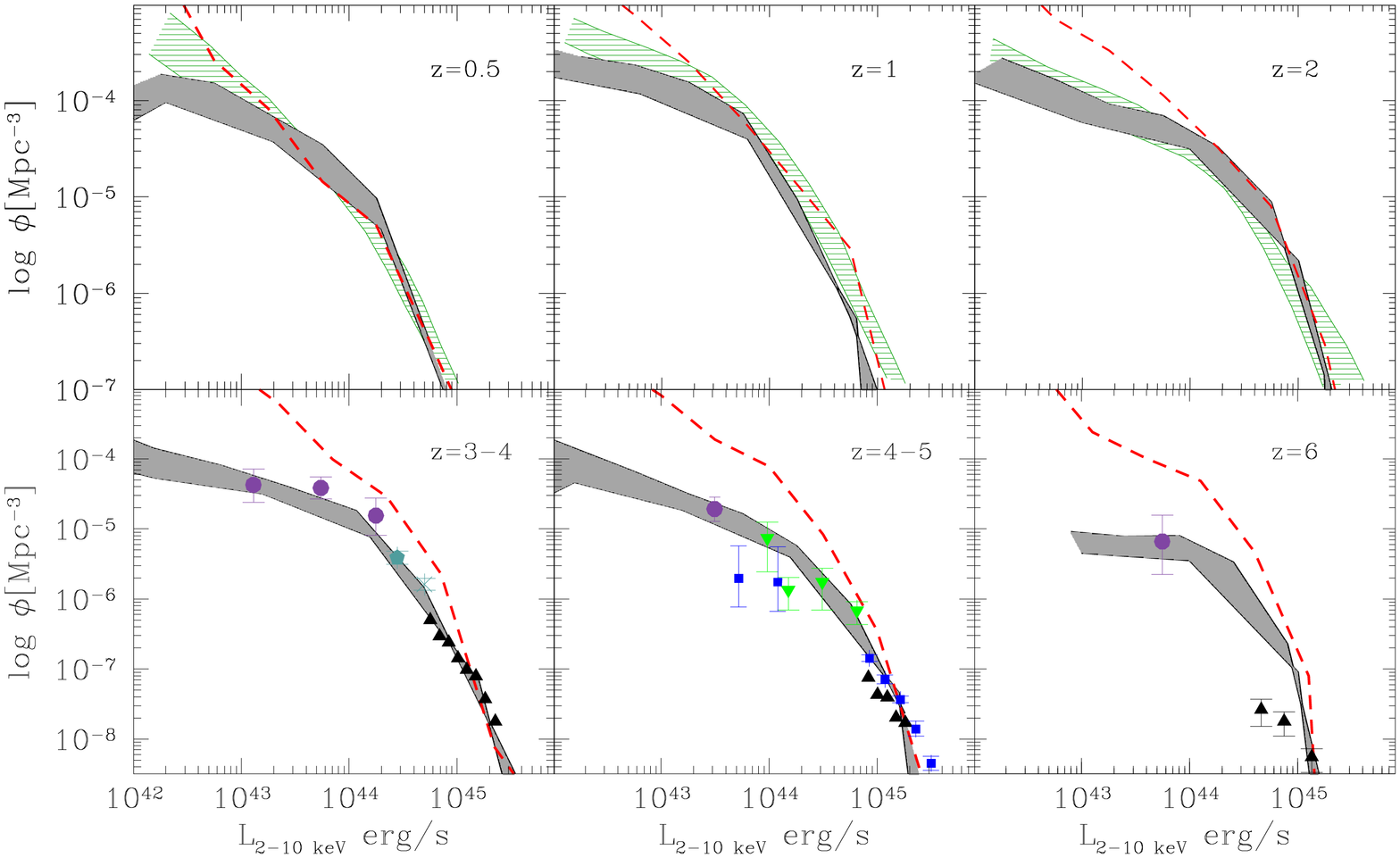}}}
\end{center}
\vspace{-1.2cm }
 {\footnotesize 
Fig. 1. - The evolution of  the AGN luminosity function $\phi=dn/\d log L$ (in the X-ray band 2-10 keV) predicted in WDM cosmology (the grey region) is compared with that corresponding to a CDM cosmology (dashed red line) and with the results of different observations; 
the upper envelope of the grey region correspond to WDM merging trees computed through direct application of the EPS theory with 
a WDM spectrum given in eq. (2) , while the lower envelope corresponds to trees where the inclusion of haloes with mass $M\leq M_{fs}$ is suppressed (see sect. 2.1).  The redshifts corresponding to the different panels are given in the labels. 
 As for the observational data (all accounting for obscured objects), the hatched regions in the low-redshift bins $z\leq 2$ bracket the observational estimates by La Franca et al. (2005), Ebrero et al. (2009) and Aird et al. (2010)
and thus take into account systematic uncertainties of the different determinations.  At higher redshifts we compare with  X-ray observations of  AGN  from XMM-COSMOS  (Brusa et al. 2010, asterisk), Chandra-COSMOS (Civano et al. 2011, diamond), and  Chandra-GOODS-MUSIC/ERS (Fiore et al. 2012, solid circles). These have been complemented with measurements derived 
from UV selected samples from the GOODS (Fontanot et al. 2007, solid squares), the Sloan survey (Richards et al. 2006 and 
Jiang 2009, filled triangles) and the deep surveys analysed by Glikman et al. (2011, inverted triangles), where the rest frame 1450 {\AA} luminosities have been converted to the 2-10 keV band using the bolometric corrections in Marconi et al. (2004). 
\vspace{0.3cm}}

Thus, high-redshift $z\geq 2$ measurements of the abundance of low-luminosity $L_{2-10}\leq 10^{44}$ erg/s AGN are best suited to discriminate between the CDM and the WDM predictions. Such observations are now within the reach of the recent deep searches. In particular, Fiore et al. (2012) 
exploited the combination of the multicolor catalogues MUSIC and ERS of the GOODS south field (Grazian et al. 2006, 2011) with 
4 Ms Chandra images in the 2-10 KeV X-ray band to obtain the deepest measurements of low-luminosity AGN. 
They selected faint AGN activity among the numerous Lyman break galaxies in a wide redshift interval $3<z<7$ putting interesting constraints on the faint end of the luminosity function at $z=3-4, 4-5$, and $z>5.8$, that we show in the bottom panels of fig. 1. Note that such points account for obscured objects, 
including the Compton-thick fraction observed in the sample ($\sim 20\%$ for $z\gtrsim 3$); Fiore et al. (2012) estimated the uncertainty on the observed luminosity functions due to possibly undetected (or unaccounted-for) Compton thick sources to be small ($\lesssim 15\%$). Thus,  
the  X-ray detections make the sample  more complete with respect to the optical surveys and this is reflected in the higher densities attained by the Fiore et al. (2012) luminosity functions compared to extrapolations of their optical counterparts to low luminosities; nevertheless, for $L_{2-10}\leq 10^{44}$ erg/s, such number densities are  overestimated by our fiducial CDM model by factors up to ten (in the lowest luminosity bins), while they are well matched by the WDM models.
Thus, the indications from such deep, high-redshift observations is that the WDM scenario that provides a viable solution to the CDM 
over-prediction of faint galaxies (see Paper I) could also provide an interesting framework for the evolution of the AGN population, at least in interaction-driven scenarios for the AGN triggering. As we shall show in Sect. 4, such a conclusion is quite robust with respect to the choice of the Supernovae feedback parameters, the main source of uncertainty at the faint end of the AGN luminosity function in semi-analytic models. 

Finally, we note that the correlation between the galactic and the AGN properties is not strongly affected by the shape of the DM power spectrum on small scales. This is shown in fig. 2, where we compare the local Black Hole-Stellar mass relations predicted in the CDM and WDM scenarios with availble data (details on the comparison are given in Lamastra et al. 2010, where we presented the CDM predictions). The only appreciable difference is the smaller scatter (for small values of $M_{BH}$) in the WDM case, due to the smaller number of interactions with low-mass galaxies, largely affecting the accretion history of low-mass BHs. 
The predicted evolution of the BH-stellar mass relation is also weakly dependent on the adopted power spectrum: this is illustrated by the paths showing the growth of BH mass relative to the stellar mass for largest BHs in our Monte Carlo simulation. As predicted for CDM in our previous work (see Lamastra et al. 2010), the local relation is reached - for massive objects- through paths passing {\it above} the local relation, indicating a faster growth of BH masses compared to stellar masses in the earliest phase of galaxy evolution, in both the CDM and the WDM case. 

\begin{center}
\vspace{-0.4cm}
\scalebox{0.36}[0.36]{\rotatebox{0}{\includegraphics{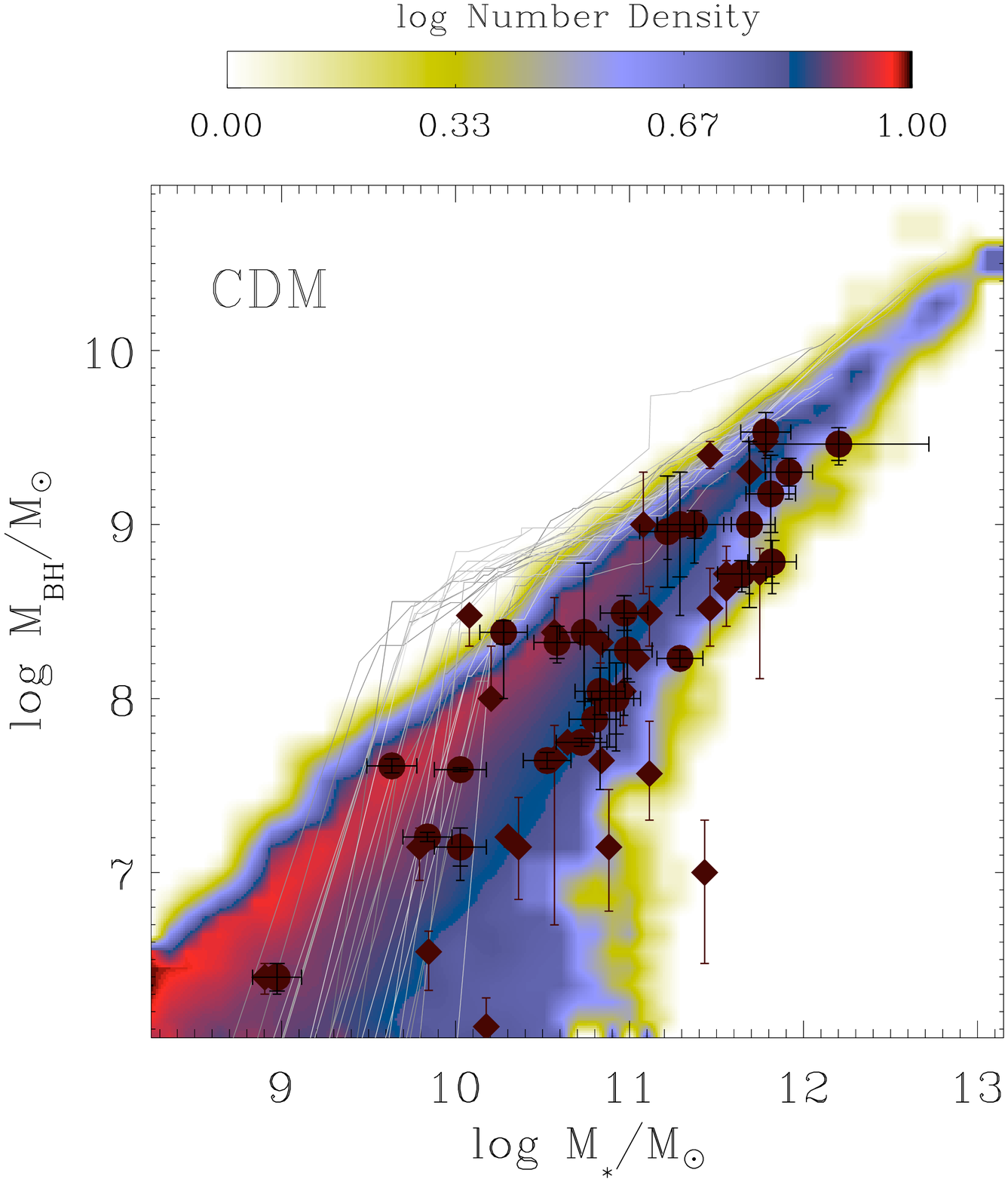}}}
\scalebox{0.36}[0.36]{\rotatebox{0}{\includegraphics{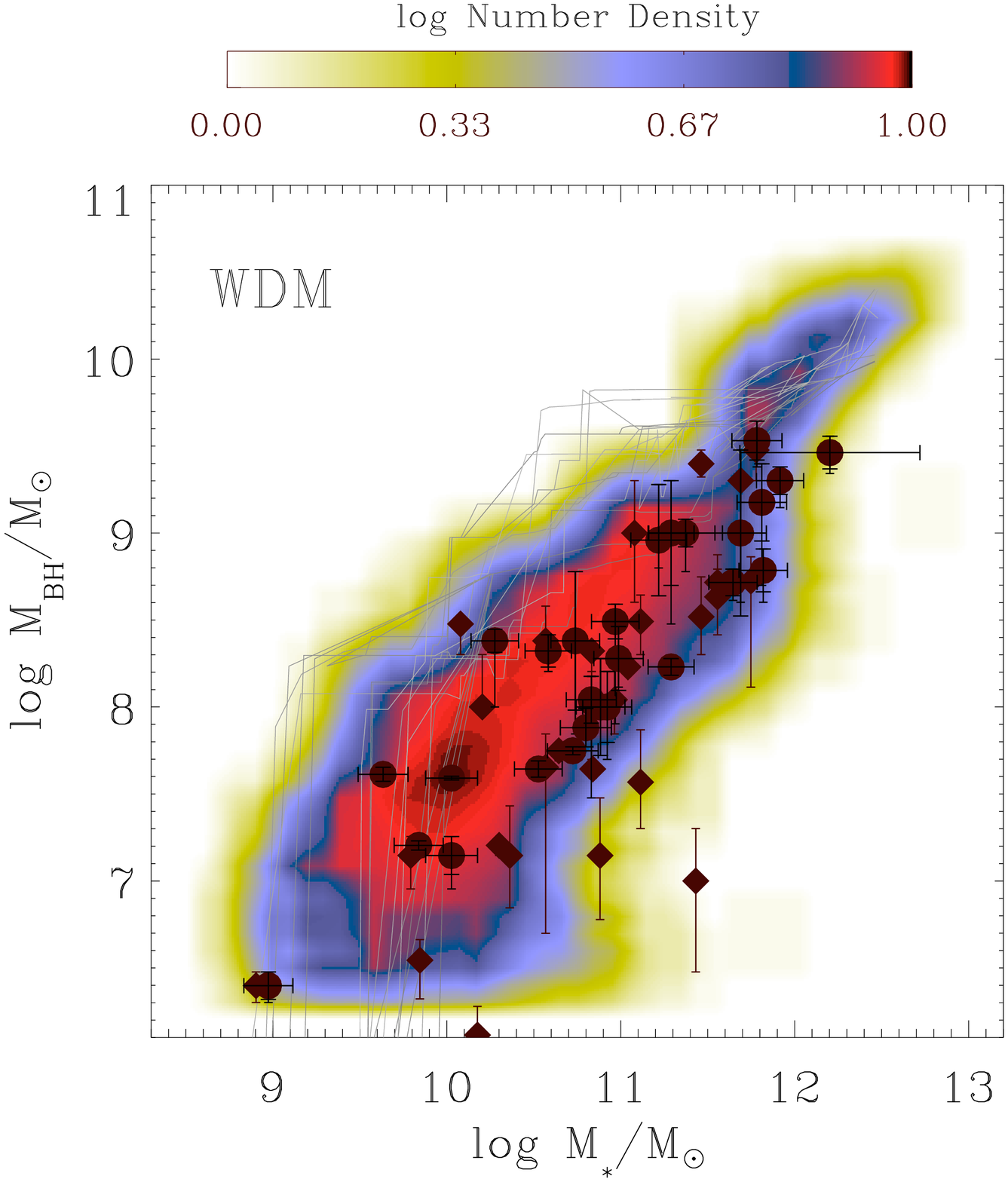}}}
\end{center}
\vspace{-0.5cm }
 {\footnotesize 
Fig. 2. -The local predicted $M_{BH}-M_*$ relation for the CDM  (left panel) and the WDM (right panel) cosmology is compared with data by H\"aring and Rix (2004, diamonds), and Marconi and Hunt (2003, squares); the color code represents the logarithm of the space density of BHs in a given $M_{BH}-M_*$, normalized to the maximum value, as indicated by the upper color bar. We also show some of the paths in the $M_{BH}(t)-M_*(t)$ plane followed, during their evolution, by BHs (and by their host galaxies) reaching a final  mass of $M_{BH}(z=0)\geq 10^{10}\,M_{\odot}$.}
\vspace{0.2cm}

\section{Discussion. The Role of Galaxy Feedback}
In the previous section we have shown that the recent observational estimates of the high-redshift AGN luminosity functions are matched by the model predictions in WDM cosmology, due to the suppression of the {\it number density} of low mass galaxies at high redshifts . It is interesting to investigate whether a similar agreement can be achieved through a proper {\it luminosity evolution} within the standard CDM scenario. This would require  a  suppression the AGN luminosities for $L_X\lesssim 10^{44}$ erg/s at redshift $z\gtrsim 3$.  Here we investigate whether increasing the efficiency and the mass dependence of the stellar feedback can help in bringing the CDM predictions in closer agreement with present data on the AGN abundance at high redshifts. 

To this aim, we modified our fiducial CDM model by maximizing the feedback efficiency at small galaxy masses. Our {\it maximal feedback} model differs from our fiducial case in three respects: i) the (inverse) scaling of the galaxy feedback on the circular velocity  is increased from $\dot m_{hot}\propto v_c^{-2}$ (see sect. 2.2) to $\dot m_{hot}\propto v_c^{-4}$; ii) the Supernovae efficiency is increased to $\epsilon=0.03$; iii) an artificial cutoff for gas cooling is assumed for galaxy haloes with circular velocity larger than $250$ km/s; this mimics the effect of the AGN radio mode implemented in several SAMs to fit the bright end of the local luminosity functions (the chosen value for the cutoff circular velocity corresponds to that resulting from the analysis of Fontanot et al. 2011).  Although such a point constitutes a critical issue in galaxy formation models, it has a minor influence on the evolution of the low-mass end of the galaxy and AGN luminosity functions

Such a {\it high feedback} model is similar to other semi-analytic models, like that by Guo et al. (2011) which adopt a similar scaling for the SN feedback. In fact, it provides an excellent fit to the low-redshift stellar mass distribution and luminosity function (see fig. 3a,b), similar to that provided by our WDM model; even though it is well known that such high-feedback models in CDM cosmology yield an exceeding fraction of faint red galaxies at low redshift (see also Guo et al. 2011; Lo Faro et al. 2009), we shall investigate such a case to probe the effects of changing the galaxy feedback on the AGN evolution, and to bracket the range of CDM predictions corresponding to different SAMs in the literature (within the assumption of interaction-triggered AGN fueling). 

The evolution of the AGN luminosity functions in the above "maximal-feedback" model is shown in the bottom panels of fig. 3, where we also show for comparison the results for the WDM (solid line) and the fiducial CDM model (dotted).  Note that, while at low redshifts  the 
CDM maximal feedback model provides an excellent match to the data, at redshifts $z\gtrsim 3.5$ the implementation of a maximal feedback does not seem able to properly match the observed abundance of low-luminosity AGN (those in the faintest bin $L_{2-10}\leq 5\,10^{43}$ erg/s). This is due to a twofold reason: i) the inner density of DM haloes rapidly grows with redshifts, so that for $z\gtrsim 2$ the escape velocity of DM haloes $\propto \rho^{1/3}$ remains large even for low-mass galaxies, and  this, in turn, suppresses the feedback efficiency in removing gas from the cold phase. This effect has already been found and discussed by, e.g.,  Lo Faro et al. (2009), who found that SAMs in CDM cosmology over-predict the slope of the  low-mass end of the stellar mass distribution at high-redshifts $z\gtrsim 2$ even in the presence of a large feedback efficiency; this effect is also responsible for the over-prediction of low-mass galaxies at $z\gtrsim 1$ in the SAM by Guo et al. (2011) based on the Millenium simulation.
ii) in interaction-driven models for the AGN feeding, faint AGN at high redshifts are associated to minor-mergers, rather than being produced by gas-poor galaxies; to reduce the abundance of faint AGN with respect to the fiducial CDM case, the suppression of the number density of low-mass galaxies (and hence of minor mergers) resulting in WDM cosmology is more efficient than reducing the cold gas fractions by increasing the stellar feedback. 

Thus, at high redshifts $z\gtrsim 2$  the CDM over-prediction of  low-mass galaxies yield by CDM models (see Lo Faro et al. 2009; Guo et al. 2011; Somerville et al. 2008) seems to have a counterpart in an excess of the number of predicted faint AGN even assuming a maximal feedback efficiency; this makes the number density effects of WDM scenario an attractive alternative for all models aiming at providing a unified solution to the different aspects of the CDM over-abundance problem. 
\begin{center}
\vspace{-0.5cm}
\scalebox{0.6}[0.6]{\rotatebox{-90}{\includegraphics{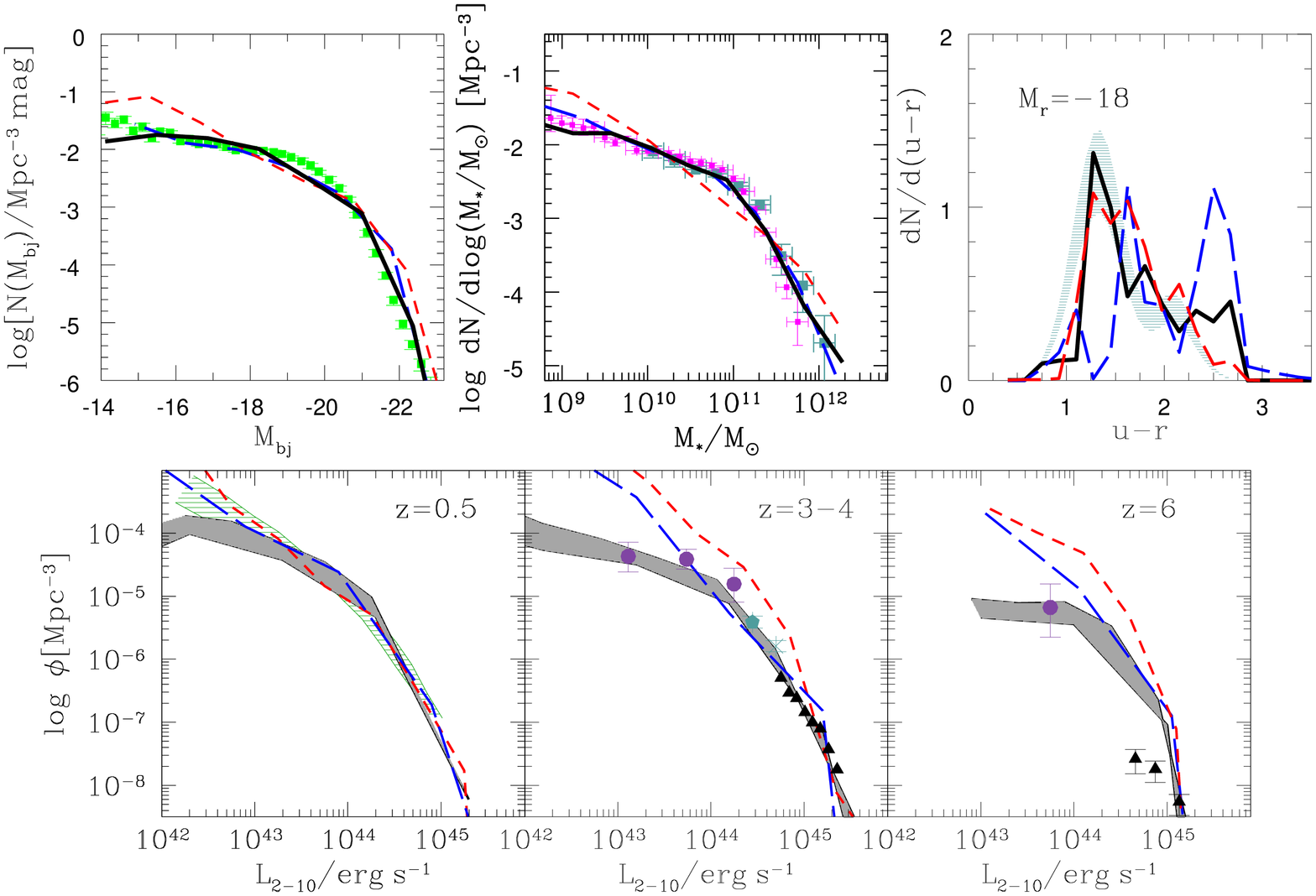}}}
\end{center}
\vspace{-0.2cm }
 {\footnotesize 
Fig. 3 - {\it Top Left}: The local galaxy luminosity function ($b_j$-band) in the WDM model (solid line) compared to our maximal feedback CDM model (see text, long dashed blue line); we also show the corresponding prediction for our fiducial CDM model (short-dashed red line): the data points are from the 6dF Galaxy Survey (Jones et al. 2006); for the sake of readability, only the WDM model derived from direct extension of EPS (see sect. 2.1) is shown. 
 {\it Top Central}: 
The local stellar mass distribution: the different models correspond to line types as in previous panel. Data points are from Drory et al. (2004, squares) and Fontana et al. (2006, circles). {\it Top Right}: 
The $u-r$ color distribution of faint ($M_r=-18$) galaxies in the different models (line types as in previous panels) is compared with the Sloan data  (from Baldry et al. 2004, hatched area).\newline
{\it Bottom Panels}: The evolution of  the AGN luminosity function (in the X-ray band 2-10 keV) predicted in WDM models (grey region, as in fig. 1) is compared with that corresponding to our {\it maximal feedback} CDM model (long dashed blue line) to our fiducial CDM model (short-dashed red line) and to the results of different observations. The redshifts corresponding to the different panels are given in the labels. Data hatched region and data points as in fig. 1.}
\vspace{0.3cm}

\section{Conclusions}

We have shown that the recent measurements of the abundance of low-luminosity  ($L_X\leq 10^{44}$ erg/s) AGN at high redshifts $z\geq 4$ can provide severe constraints to models connecting the AGN evolution to galaxy formation through the effects of galaxy interactions. 
In particular, the long-standing problem of CDM over-prediction of low-mass galaxies at such high redshifts reflects into a similar excess of low-luminosity AGN. We have shown that the suppression of the number of low-mass galaxies obtained in a WDM model (with WDM particles of mass $m_X\approx 1$ keV) provides a viable solution to both the galactic (Menci et al. 2012) and the AGN (this paper) excess. We have also shown that solutions to such excess based on increased Supernovae feedback are  less efficient than the WDM suppression of low-mass galaxies, at least in interaction-driven models for the triggering of AGN; in fact, in such  models, faint AGN at high redshifts are associated to minor-mergers, rather than being produced by gas-poor galaxies; to reduce the abundance of faint AGN with respect to the fiducial CDM case, the suppression of the number density of low-mass galaxies (and hence of minor mergers) resulting in WDM cosmology is more efficient than reducing the cold gas fractions by increasing the stellar feedback. Note however that, when the WDM merging trees are computed with progenitor distributions and halo mass functions strongly suppressed below the free-streaming scale (as suggested by Benson et al. 2013), the low-redshift  AGN luminosity function under-predicts the observations at low luminosities $L_{2-10}\leq 10^{43}$ erg $s^{-1}$ (see fig. 1). 

While the strongest constraints to distinguish between WDM and CDM models are naturally given by the faint ($L_X\leq 10^{44}$ erg/s) end of the  AGN luminosity function at $z\geq 4$, faint AGN at high-z is of course the region of the AGN parameter space most
difficult to probe. The measurements of the faint end of the AGN luminosity function in Figs. 1 and 3 are reasonably good at z=3-4, but they 
are rather loose at z$>4$. In particular, as noted by Fiore et al. (2012) the point at z=6 should be regarded as an upper limit to the
AGN density rather than a determination, being based on just two galaxies, detected in the small ERS area within the CDFS region, with
rather loosely constrained photometric redshift. As of today we do not have a single AGN at z$>6$ with a robust spectroscopic redshift. This
poor situation may improve in the next few years.  At the flux limits reached by the deepest Chandra exposure (4 Mseconds) there
are several hundreds AGN/deg$^2$ for z$>4$, a number $\sim 100$  AGN/deg$^2$ for z$>5$, and $< 100$ AGN/deg$^2$ for z$>5.8$. 
It is clear that to obtain a more robust
demography of the z$>$6 AGN a search in a much wider area, such as the CANDELS area (Grogin et al. 2011; Koekemoer et al. 2011), is mandatory, and requires
spectroscopic confirmation of the X-ray emitting, candidate z$>6$ galaxies. The CANDELS deep and wide surveys cover a total of 130
arcmin$^2$ and 670 arcmin$^2$ to a depth of H=27.8 and H$\sim26.5$ respectively, about 3 times and 12 times the ERS area. The two
candidate z$>6$ ERS galaxies detected by Chandra in the ERS field are faint, H=26.6 and H=27 sources. The GOODS source with z$>7$ in the Luo
et al. (2010) catalog has H=27.6. In summary, we expect a number $<5$ AGN in the CANDELS deep survey for z$>6$ and less than $20$ AGN for z$>6$  in the CANDELS wide survey. We note that a fraction of these sources will be at the limit, or below, the H band sensitivity threshold of the
wide survey. As of today, Chandra has spent of the order of 8 Mseconds on the CANDELS fields, most of them on the CANDELS deep fields.  To
reach the sensitivity to detect the faint z$>6$ AGN in the wide area, additional 5-6 Mseconds are needed.  This is within reach of the
Chandra observatory in the next few years. 

Note that, our results do not exclude that the CDM predictions for the number low-luminosity  AGN at high redshifts can be reconciled 
with observations provided: i) less luminous AGNs are driven by secular processes rather than by galaxy interactions; ii) 
additional baryonic physics is involved in the Black Hole accretion at high redshifts (especially for low-luminosity AGNs); 
iii) heavy initial BH seeds. 
Such processes could provide the negative AGN luminosity evolution needed to balance the AGN excess of standard CDM models. 
Indeed, some of the above conditions have been explored in recent semi-analytic models. As for the first possibility (point i) above), 
both Fanidakis et al. (2012) and Hirschmann et al. (2012) also include disk instabilities as an additional trigger for BH accretion, with the former paper also 
including accretion at low rates from hot halos. While such models provide excellent fits to the 
AGN luminosity functions at $z\leq 4$, at higher redshifts $z\gtrsim 4$ they predict different abundances for low-luminosity AGNs; while the 
predictions shown in Fanidakis et al. (2012) are consistent with observations, the Hirschmann et al. (2012) model overestimates the abundance of 
faint AGNs. Here the critical issue is 
the luminosity range where the different accretion modes take over; in this respect, important hints to improve modelling will be provided by observations 
relating the AGN luminosity with galaxy mergers (see Treister et al. 2012) and with giant clumps associated to disk instabilities 
(see Bournaud et al. 2012) as well as by aimed N-body simulations (see, e.g., Bournaud et al. 2011). 

As for the second and third possibilities (points ii) and iii) above), they are still open; indeed, Hirschmann et al. (2012) have shown that 
assuming a varying sub-Eddington limit ( that depends on the cold gas fraction), and heavy BH seeds with a tuned cutoff for the host halo mass can resolve the mismatch. 
However, while varying sub-Eddington limit onto low-mass BHs at high redshifts and heavy BH seeds could solve the CDM excess of predicted low-luminosity AGNs, they would leave unchanged its over-prediction of low-mass/faint galaxies at high redshifts $z\gtrsim 3$, which should find a different solution. Indeed, our results strongly indicate that, when seeking for a unified a solution to the CDM over prediction of low-luminosity AGN  and of low-mass galaxies at $z\geq 3$, the WDM cosmology represents an appealing solution.  

\acknowledgments We acknowledge grants from INAF and from ASI-INAF contract I/009/10/0.


\begin{thebibliography}{}
\bibitem{}Abadi, M. G., Navarro, J. F., Steinmetz, M.  Eke, V. R. 2003, ApJ 591, 499
\bibitem{}Bahcall, J. N., Kirhakos, S., Saxe, D. H., Schneider, D. P. 1997, ApJ, 479, 642
\bibitem{}Baldry, I.K., Glazebrook, K., Brinkmann, J., Zeljko, I., Lupton, 
R.H., Nichol, R.C., Szalay, A.S. 2004, ApJ, 600, 681 
\bibitem{}Bardeen J. M., Bond J. R., Kaiser N., Szalay A. S., 1986, ApJ, 304, 15 
\bibitem{}Barkana R., Haiman Z., Ostriker J. P., 2001, ApJ, 558, 482
\bibitem{}Baugh, C.M. 2006, Rep. Prog. Phys. 69, 3101
\bibitem{}Begelman, M.C. 2003, in {\it Coevolution of Black Holes and Galaxies}, Carnegie Observatories Astrophysics Series, Vol. 1,
ed. L.C. Ho (Cambridge: Cambridge University Press)
\bibitem{}Begelman, M., Volonteri, M., Rees, M. J. 2006, MNRAS, 370, 289
\bibitem{}Bennert, N., Canalizo, G., Jungwiert, B., Stockton, A., Schweizer, F., Peng, C. Y., Lacy, M. 2008, ApJ, 677, 846
\bibitem{}Benson, A.J., Frenk, C.S., Lacey, C.G., Baugh, C.M., Cole, S.  2002, MNRAS, 333, 177
\bibitem{}Benson, B.A. et al. 2013, ApJ, 763, 147
\bibitem{}Bode, P., Ostriker, J.P., Turok, N. 2001, ApJ, 556, 93
\bibitem{}Bond, J.R., Szalay, A.S., Turner, M. S. 1982, Phys. Rev. Lett. 48, 1636
\bibitem{}Bond, J.R., Cole, S., Efstathiou, G., \& Kaiser, N.,1991, ApJ, 379, 440
\bibitem{}Bower, R.G., Benson, A.J., Malbon, R., Helly, J.C., Frenk, C.S., Baugh, C.M., Cole, S., Lacey, C.G. 2006, MNRAS, 370, 645
\bibitem{}Boyle, B.J., Shanks, T., Croom, S.M., Smith, R.J., Miller, L., Loaring, N., Heymans, C. 2000, MNRAS, 317, 1014
\bibitem{}Bournaud, F. et al., 2011, ApJ, 741, L33
\bibitem{}Bournaud, F. et al., 2012, ApJ, 757, 81
\bibitem{}Brusa, M., Civano, F., Comastri, A., et al. 2010, ApJ, 716, 348
\bibitem{}Canalizo, G., \& Stockton, A. 2001, ApJ, 555, 719 
\bibitem{}Cattaneo, A., Dekel, A., Devriendt, J., Guiderdoni, B., Blaizot, J., 2006, MNRAS, 370, 1651
\bibitem{}Cavaliere, A.,  \& Vittorini, V.  2000, ApJ, 543, 599
\bibitem{}Cavaliere, A., Lapi, A., Menci, N., 2002, ApJ, 581, L1
\bibitem{}Chartas, G., Brandt, W. N., Gallagher, S. C., Garmire, G. P., 2002, ApJ, 579, 169
\bibitem{}Ciotti, L. 2009, Nature, 460, 333 
\bibitem{}Cirasuolo M., McLure R. J., Dunlop J. S., Almaini O., Foucaud S., Simpson C., 2010, MNRAS, 401, 1166
\bibitem{}Civano, F., Brusa, M., Comastri, A. et al. 2011 ApJ, 741, 91
\bibitem{}Cole, S., Aragon-Salamanca, A., Frenk, C.S., Navarro, J.F., Zepf, S.E., 1994, MNRAS, 271, 781 
\bibitem{}Cole, S., Lacey, C.G., Baugh, C.M., Frenk, C.S., 2000, MNRAS, 319, 168 
\bibitem{}Cole, S. 2005, MNRAS, 362, 505
\bibitem{}Colombi, S., Dodelson, S., Widrow, L.M. 1996, ApJ, 458, 1
\bibitem{}Croom, S.M., Richards, G.T., Shanks, T., et al. 2009, MNRAS, 399, 1755
\bibitem{}Cox, T.J. et al. 2008, MNRAS, 384, 386
\bibitem{}Croton, D.J, Springel, V., White, S.D.M., De Lucia, G.,Frenk, C.S., Gao, L., Jenkins, A., Kauffmann, G., Navarro, J.F.,
Yoshida, N., 2006, MNRAS, 365, 11
\bibitem{}de Vega, H.J., Sanchez, N.G. 2012,  Phys. Rev. D85, 3518
\bibitem{}De Lucia G., Blaizot J., 2007, MNRAS, 375, 2 
\bibitem{}De Lucia G., Boylan-Kolchin M., Benson A. J., Fontanot, F., Monaco P., 2010, MNRAS, 406, 1533
\bibitem{}Dekel, A., Silk, J. 1986, ApJ, 303, 39
\bibitem{}Di Matteo, T., Springel, V., Hernquist, L., 2005, Nature, 433, 604
\bibitem{}Drory, N. et al. 2004, ApJ, 608, 742
\bibitem{}Fan X. et al. 2001, AJ, 122, 2833
\bibitem{}Fan X. et al. 2004, ApJ, 128, 515
\bibitem{}Fanidakis, N. et al. 2012, MNRAS, 419, 2797
\bibitem{}Ferrarese, L. Merritt, D., 2000, ApJ, 539, L9
\bibitem{}Fiore, F., et al. 2012, A\&A, 537, 22
\bibitem{}Fisher, K. B., Bahcall, J. N., Kirhakos, S., Schneider, D. P. 1996, ApJ, 468, 496
\bibitem{}Fontana A. et al., 2006, A\&A, 459, 745
\bibitem{}Fontanot, F., Cristiani, S., Monaco, P., Nonino, M., Vanzella, E., Brandt, W. N., Grazian, A., \& Mao, J. 2007, A\&A, 461, 39
\bibitem{}Fontanot, F. et al. 2009, MNRAS,  397, 1776
\bibitem{}Fontanot, F. et al. 2011, MNRAS,  413, 957
\bibitem{}Gebhardt, K. et al. 2000, ApJ, 539, L13
\bibitem{}Gentile, G. Tonini, C. Salucci, P. 2007, A\&A, 467. 925
\bibitem{}Glikman, E., Djorgovski, S.G., Stern, D., Dey, A., Jannuzi, B.T., \& Lee, K.-S. 2011, \apj, 728, L26
\bibitem{}Grazian, A., Fontana, A., de Santis, C., et al. 2006, A\&A, 449, 951
\bibitem{}Grazian, A., Castellano, M., Koekemoer, A.M., et al.  2011, A\&A, 532, 33
\bibitem{}Grogin, N.A. et al. 2011, ApJS, 197, 35
\bibitem{}Guo, Q. et al. 2011, MNRAS, 413, 101
\bibitem{}H\"aring, N., Rix, H.-W. 2004, ApJ, 604, L89
\bibitem{}Hartwick, F.D.A., Shade, D., 1990, ARA{\&}A, 28, 437
\bibitem{}	Hasinger, G., Miyaji, T., Schmidt, M. 2005, A\& A 441, 417
\bibitem{}Hennawi, J. F., et al. 2006, AJ, 131, 1 
\bibitem{}Henriques, B. et al. 2011, MNRAS, 415, 3571
\bibitem{}Hirschmann, M., Somerville, R.S., Naab, T., Burkert, A. 2012, MNRAS< 426, 237
\bibitem{}Hopkins, P.F., Hernquist, L., Cox, T.J., Di Matteo, T., Robertson, B., Springel, V. 2005, ApJ, 632, 81
\bibitem{}Hopkins, P.F. Hernquist, L., Cox, T.J., Di Matteo, T., Robertson, B., Springel, V. 2006, ApJS, 163, 1
\bibitem{}Hopkins, P. F., Hernquist, L., Cox, T. J., Di Matteo, T., Martini, P., Robertson, B., \& Springel, V. 2005a, ApJ, 630, 705 
\bibitem{}Hopkins, P.F., Cox, T.J., Kereš, D., Hernquist, L., 2008, ApJS, 175, 390
\bibitem{}Hopkins, P.F., Quataert, E. 2010, MNRAS, 407, 1529
\bibitem{}Jiang, L., Fan, X., Bian, F. et al. 2009, AJ, 138, 305
\bibitem{}Jones, D.H., Peterson, B.A., Colless, M., Saunders, W. 2006, MNRAS, 396, 535
\bibitem{}Kauffmann, G., White, S.D.M., \& Guiderdoni, B., 1993, MNRAS, 264, 201 
\bibitem{}Kauffmann, G., Haehnelt, M., 2000, MNRAS, 311, 576
\bibitem{}Khlopov, M.Y., Rubin, S.G., Sakharov, A.S. 2004, Astroparticle Physics, 23, Issue 2, 265
\bibitem{}Klypin, A., Kravtsov, A.V., Valenzuela, O., Prada, F. 1999, ApJ, 522, 82
\bibitem{}Koekemoer, A. M. et al. 2011, ApJS 197, 36
\bibitem{}Kormendy, J., Richstone, D. 1995, ARA\&A, 33, 581
\bibitem{}Kormendy, J., \& Bender, R., 2009, ApJ, 691, L142
\bibitem{}Kusenko, A. 2009, Phys.Rept., 481,1
\bibitem{}Lacey, C., \& Cole, S., 1993, MNRAS, 262, 627 
\bibitem{}La Franca, F. et al. 2005, ApJ, 635, 864
\bibitem{}Lamastra, A., Menci, N., Maiolino, R. Fiore, F., Merloni, A. 2010, MNRAS, 405, 29
\bibitem{}Lapi, A., Cavaliere, A., \&  Menci, N. 2005, ApJ, 619, 60
\bibitem{}Lapi, A., Shankar, F., Mao, J., Granato, G.L., Silva, L., De Zotti, G., Danese, L. 2006, ApJ, 650, 42
\bibitem{}Lo Faro. B, Monaco, P., Vanzella, E., Fontanot, F., Silva, L., Cristiani, S. 2009, MNRAS, 399, 827
\bibitem{}Lovell, M.R. et al. 2012, MNRAS, 420, 2318
\bibitem{}Luo B., et al. 2010, \apjs , 187, 560 
\bibitem{}Madau, P., and Rees, M.J., 2001, ApJ, 551, L27
\bibitem{}Madau, P. Diemand, J., Kuhlen, M. 2008, ApJ, 679, 1260
\bibitem{}Magorrian, J. et al. 1998, AJ, 115, 2285
\bibitem{}Marchesini, D. et al. 2009, ApJ, 701, 1765
\bibitem{}Marconi, A., Hunt, L.K., 2003, ApJ, 589, L21
\bibitem{}Marconi, A., Risaliti, G., Gilli, R., Hunt, L.K.,Maiolino, R., Salvati, M. 2004, MNRAS, 351, 169
\bibitem{}Marulli, F., Bonoli, S., Branchini, E., Moscardini, L., Springel, V. 2008, MNRAS, 385, 1846
\bibitem{}Mateo, M. 1998, ARA\&A, 36, 435
\bibitem{}McGaugh, S.S., de Blok, W.J.G., Schombert, J.M., Kuzio de Naray, R., Kim, J.H., 2007, ApJ, 659, 149
\bibitem{}Menci, N., Cavaliere, A., Fontana, A., Giallongo, E., Poli, F., 2002, ApJ, 578, 18
\bibitem{}Menci, N., Cavaliere, A., Fontana, A., Giallongo, E., Poli, F., Vittorini, V. 2003, ApJ, 587, L63
\bibitem{}Menci, N., Cavaliere, A., Fontana, A., Giallongo, E., \& Poli, F. 2004,ApJ, 604, 12
\bibitem{}Menci, N., Fontana, A., Giallongo, E.,  Salimbeni, S. 2005, ApJ,  632, 49
\bibitem{}Menci, N., Fontana, A., Giallongo, E., Grazian, A.,Salimbeni, S. 2006, ApJ,  647, 753
\bibitem{}Menci, N., Fiore, F., Puccetti, S., Cavaliere, A. 2008, ApJ, 686, 219
\bibitem{}Menci, N., Fiore, F., Lamastra, A. 2012, MNRAS, 421, 2384 (Paper I)
\bibitem{}Monaco, P., Fontanot, F., Taffoni, G., 2007, MNRAS, 375, 1189
\bibitem{}Moore, B. et al.  1999, MNRAS, 310, 1147
\bibitem{}Myers, A. D., Richards, G. T., Brunner, R. J., Schneider, D. P., Strand, N. E., Hall, P. B., Blomquist, J. A., York, D. G. 2008, ApJ, 678, 635 
\bibitem{}Natarajan, P. 1999, ApJ, 512, 105
\bibitem{}Pagels, H., Primack, J. 1982, Phys. Rev. Lett., 48, 223
\bibitem{}Peebles, P.J.E. 1982, ApJ, 258, 415
\bibitem{}Peebles, P.J.E. 1993, {\it Principles of Physical Cosmology} (Princeton: Princeton Univ. Press)  
\bibitem{}Pounds, K., King, A.R., Page, K.L., O'Brien, P.T 2003, MNRAS, 346, 1025
\bibitem{}Prokhorov, D. \& Silk, J. 2010, 725, 131
\bibitem{}Reed, D., Governato, F., Verde, L., Gardner, J.,Quinn, T., Stadel, J., Merritt, David; Lake, G. 2005, MNRAS, 357, 82
\bibitem{}Richards, G.T. et al. 2006, ApJ, 131, 2766
\bibitem{}Salimbeni, S. et a. 2008, A\&A, 477, 763
\bibitem{}Sanders, D. B., \& Mirabel, I. F. 1996, ARA\&A, 34, 749
\bibitem{}Santini, P., et al. 2011, A\&A, 540, 109
\bibitem{}Saslaw, W.C., 1985, {\it  Gravitational Physics of Stellar and Galactic Systems} (Cambridge: Cambridge Univ. Press) 
\bibitem{}Serber, W., Bahcall, N., M\'enard, B., Richards, G. 2006, ApJ, 643, 68 
\bibitem{}Shankar F., \& Mathur S. 2007, ApJ, 660, 1051 
\bibitem{}Shankar F., Bernardi, M. \& Haiman, Z. 2009, ApJ, 694, 867 
\bibitem{}Shankar F., Crocce M., Miralda-Escud´e J., Fosalba P., Weinberg D. H. 2010, ApJ, 718, 231
\bibitem{}Shankar F., Weinberg D. H., \& Miralda-Escud´e J. 2009, ApJ, 690, 20
\bibitem{}Shen, Y. 2009, ApJ, 704, 89
\bibitem{}Sheth, R.K., Mo, H.J., Tormen, G. 2001, MNRAS, 323, 1
\bibitem{}Shin, M. S., Ostriker, J. P., Ciotti, L. 2010, ApJ, 711, 268
\bibitem{}Sijacki, D., Springel, V., Di Matteo, T., Hernquist, L. 2007, MNRAS, 380, 877
\bibitem{}Smith, R. E., Markovic, K.  2011, Phys. Rev. D84, 3507
\bibitem{}Somerville, R.S., \& Primack, J.R., 1999, MNRAS, 310, 1087 
\bibitem{}Somerville, R. 2002, ApJ, 572, 23
\bibitem{}Somerville, R.S., Hopkins, P.F., Cox, T.J., Robertson, B.E., Hernquist, L.  2008, MNRAS, 391, 481
\bibitem{}Spergel, D.N. et al. 2006, ApJ,  in  press (astro-ph/0603449)
\bibitem{}Springel, V. 2005, Nature, 435, 629
\bibitem{}Strand, N. E., Brunner, R. J., Myers, A. D. 2008, ApJ, 688, 180 
\bibitem{}Treister, E., Schawinski, K., Urry, C.M., and Simmons, B.D. 2012, ApJ, 758, L39
\bibitem{}Ueda, Y., Akiyama, M., Ohta, K., Miyaji, T. 2003, ApJ, 598, 886
\bibitem{}Viel, M., Legourgues, J., Haenhelt, M.G., Matarrese, S. Riotto, A. 2005, Phys. Rev. D, 71, 063534
\bibitem{}Viel, M., Becker, G.D., Bolton, J.S., Haenhelt, M.G., Rauch, M., Sargent, W.L.W. 2008, Phys. Rev. D, 100, 041304
\bibitem{}Wang, J., White, S.D.M. 2007, MNRAS, 380, 93
\bibitem{}Wyithe, J.S.B., Loeb, A. 2003, ApJ, 586, 693
\bibitem{}Yu, Q., \& Tremaine, S., 2002, MNRAS, 335, 965
\end{thebibliography}
\end{document}